\documentclass[twocolumn,prb,superscriptaddress,showpacs,preprintnumbers]{revtex4}

\usepackage{amsmath,dcolumn,graphicx}

\begin{document}

%\draft

\title{Possible mechanism for glass-like thermal conductivities in crystals with 
off-center atoms}

\author{F. Bridges} \affiliation{Physics Department, University of California, 
Santa Cruz, California 95064, USA}
\author{L. Downward} \affiliation{Physics Department, University of California, 
Santa Cruz, California 95064, USA}

\date{Draft: \today}

\begin{abstract}

In the filled Ga/Ge clathrate, Eu and Sr are off-center in site 2 but Ba is
on-center.  All three filler atoms (Ba,Eu,Sr) have low temperature Einstein
modes; yet only for the Eu and Sr systems is there a large dip in the thermal
conductivity, attributed to the Einstein modes.  No dip is observed for Ba.
Here we argue that it is the {\it off-center displacement} that is crucial for
understanding this unexplained difference in behavior.  It enhances the
coupling between the ``rattler" motion and the lattice phonons for the Eu and
Sr systems, and turns on/off another scattering mechanism (for 1K $<$ T $<$
20K) produced by the presence/absence of off-center sites. The random
occupation of different off-center sites produces a high density of
symmetry-breaking defects which scatters phonons.  It may also be important for
improving our understanding of other glassy systems.

\end{abstract}

%\pacs{65.40.-b,65.60.+a,63.20.-e 61.10.Ht}

%61.10.Ht X-ray absorption spectroscopy: EXAFS, NEXAFS, XANES, etc.
%65.40.-b Thermal properties of crystalline solids 
%65.60.+a Thermal properties of amorphous solids and glasses: 
%63.20.-e Phonons in crystal lattices 
\maketitle

%\section{Introduction}

An important, universal, characteristic of glass-like systems is a low thermal
conductivity $\kappa$; $\kappa$(T) varies roughly as T$^2$ for T $<$ 1K, forms
a plateau/dip region somewhere in the range 5-40K, and then increases slowly at
higher temperatures.  The low T behavior ($<$ 1K) is well understood, and is
attributed to a broad distribution of tunneling centers (or two level systems),
both in glasses and in some disordered crystals that exhibit glass-like
behavior such as (KBr)$_{1-x}$:(KCN)$_x$\cite{Grannan88},
while at high temperatures, Rayleigh scattering becomes important. However the
intermediate plateau/dip region is less well understood.  Several explanations
for the plateau have been proposed, but the model of Grannan {\it
etal}\cite{Grannan88} for the mixed KBr:KCN system has an important
general mechanism and appears to be applicable for both glassy and disordered
crystalline materials.  In their model, the plateau/dip is produced by a nearly
localized mode [the libration modes of the CN ion about its center of mass, at
THz frequencies] that resonantly scatters phonons very effectively. Similar
models have been applied to glassy selenium\cite{Bermejo94} and disordered
garnets\cite{Giesting02}.  The same approach has also been used to describe the
glass-like thermal conductivity in the off-center Eu and Sr filled
clathrates\cite{Nolas00}; in this case a ``rattler" mode with a low Einstein
temperature in the range 30-150K (i.e.  at THz frequencies) plays the role of
the nearly localized mode.  However this model fails in explaining the
difference in behavior between the Ba and Eu/Sr filled clathrates - no dip is
observed for the on-center Ba system.  It also cannot explain the $\kappa$ $\propto$ T
dependence observed from roughly 1-10K\cite{Sales01}. Here we will argue that
it is the off-center displacement of Eu and Sr that leads to the plateau/dip
region in these systems.

In several compounds with large unit cells (skutterudites and type I
clathrates), large cages or voids occur in the structure which can be
``filled" with several types of atoms. This dramatically alters their physical
properties\cite{Brown80,Stetson91,Evers94,Danebrock96}.  When the ``filler" ion
is considerably smaller than the void, it is loosely bound and can ``rattle"
around - hence its name. The weak binding leads to a nearly local
mode\cite{Dong01} described by a low Einstein temperature.  A low thermal
conductivity has been found in a number of such
systems\cite{Nolas99,Nolas00,Sales01} and a glass-like model for $\kappa$(T),
used for the Eu and Sr clathrates\cite{Nolas00}.  Note that a very low value of
$\kappa$ is crucial for thermoelectric applications because the figure of merit
ZT = TS$^2$$\sigma_e$/$\kappa$ (S is the Seebeck coefficient, $\sigma_e$, the
electrical conductivity) can be significantly increased. Hence understanding
the mechanisms that lead to a small $\kappa$, and particularly their ranges of
validity, is very important.

\begin{figure}[t]
\vbox{
\center{\includegraphics[width=2.8in,clip]{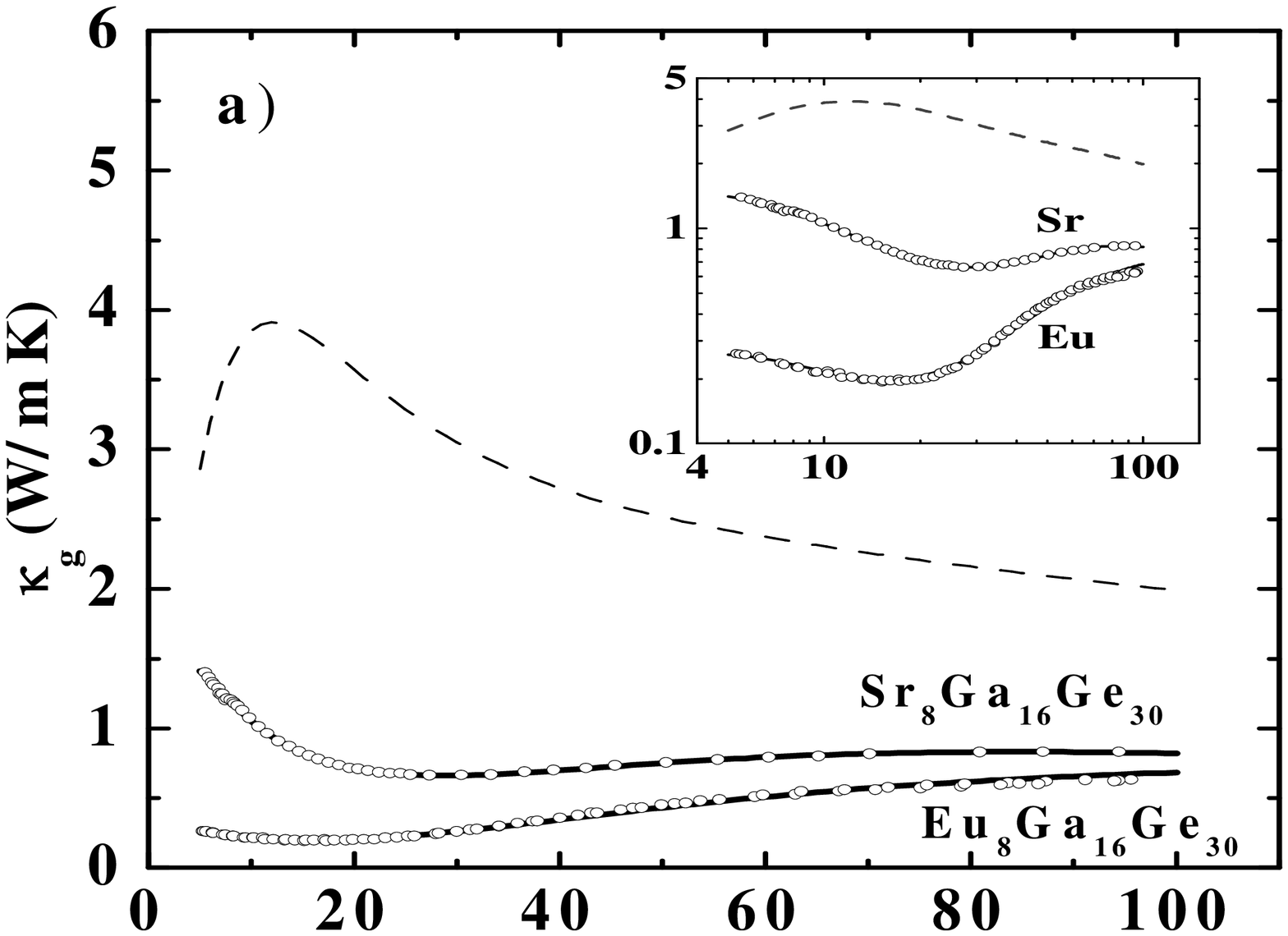}}
\center{\includegraphics[width=2.8in,clip]{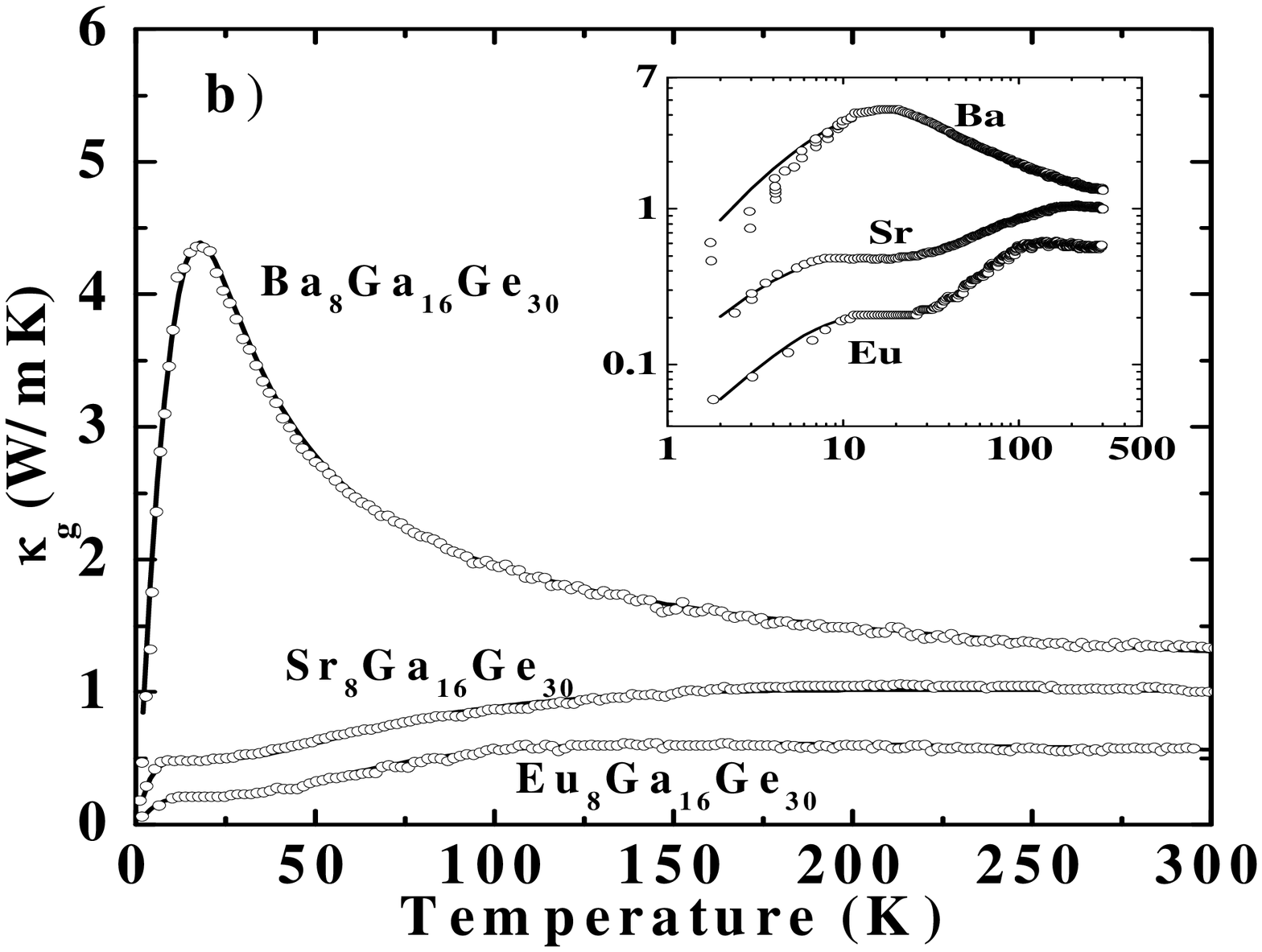}}
}
\caption{a) The lattice thermal conductivity for Eu$_8$Ga$_{16}$Ge$_{30}$ and
Sr$_8$Ga$_{16}$Ge$_{30}$ from ref \onlinecite{Nolas00}. Solid lines  -- the
new fits discussed below; dashed line -- the calculated result for
Sr$_8$Ga$_{16}$Ge$_{30}$ if the resonant scattering contribution were reduced
by a factor of 10. It shows a dependence similar to that of
Ba$_8$Ga$_{16}$Ge$_{30}$ in b. The inset shows a log-log plot. b) Similar data
for the Eu Sr, and Ba clathrates from ref \onlinecite{Sales01}. Solid lines
show the fits; for Eu and Sr the same model is used as for a) but G and the C$_i$ are
larger; for Ba, the C$_i$ are small and a finite sample size and an Umklapp 
term are needed at low T. }
\label{nolas_sales}
\end{figure}

Here we consider $\kappa$(T) for the (type I)  Ga/Ge clathrates whose structure is
formed of two cages with internal voids large enough to house another atom
(sites 1 and 2); atoms placed in them, such as Eu, are referred to as Eu1 and
Eu2. In the larger site 2 cage, Eu2, Sr2, and Ba2 all form rattlers with quite
low Einstein temperatures\cite{Sales01}; however Ba2 is on-center while both
Eu2 and Sr2 move
off-center\cite{Nolas98a,Nolas01,Chakoumakos00,Chakoumakos01,Zhang02,Bridges03}.

In Fig. \ref{nolas_sales}a, we re-plot the thermal conductivity data of Nolas
{\it etal}\cite{Nolas00}; $\kappa$ is small for Eu and Sr, and there is a well
defined dip associated with the Einstein temperatures of the
rattlers\cite{Nolas00}.  Nolas {\it etal} model the low thermal conductivity
below 20K for the Eu and Sr compounds using a broad distribution of off-center
tunneling states. At slightly higher temperatures, they used two resonant
scattering terms (two Einstein temperatures) to explain the dip (near 25K),
plus a  Rayleigh scattering term which dominates at high T. Using this sum of
terms, they were able to fit $\kappa$(T) for both Eu$_8$Ga$_{16}$Ge$_{30}$ and
Sr$_8$Ga$_{16}$Ge$_{30}$ over the range 5- 100K. However a problem arises in
using a tunneling model with tunneling splittings up to $\sim$ 15K;- first the
inferred distribution of tunneling states from ultrasonic measurements is much
narrower than in a glass\cite{Keppens00}, and second (more importantly),
very recent Mossbauer measurements indicate that the tunneling frequency is at
most 0.44 Ghz - i.e. $\sim$ 0.02K\cite{Hermann04}. If one cuts off the
tunneling density of states with a step function even at 1K, then $\kappa$
rises too rapidly below 15K.

Similar experimental results for $\kappa$(T) of these filled clathrates have
also been observed by Sales {\it etal}\cite{Sales01} (See Fig.
\ref{nolas_sales}b); however, at low T, $\kappa$ is considerably lower and the
dip is less pronounced. More importantly, for the Eu and Sr data, $\kappa$(T)
decreases nearly linearly with T below 10K, which can't be fit with the above
model. 

The T-dependence for Ba$_8$Ga$_{16}$Ge$_{30}$ in which Ba2 is {\it
on-center}\cite{Sales01}, is more perplexing (See Fig.  \ref{nolas_sales}b).
For this system, $\kappa$ is much higher overall and there is no clear dip in
$\kappa$(T), although the Einstein temperatures are only slightly larger than
those of Eu and Sr\cite{Sales01}; thus the rattler-phonon coupling must be
greatly reduced\cite{Sales01b}.  Also, the T-dependence below 10-12K is much
faster ($\kappa$ $\sim$ T$^3$) for the Ba clathrate.  These results raise two
important issues - what determines the strength of the coupling between the
rattlers and the phonons and how can one understand the T-dependence of
$\kappa$ from 1-15K?  We propose another mechanism to explain this
temperature range; it may also contribute to $\kappa$ in other
glassy systems.

The off-center displacement appears to play two additional, 
crucial roles in increasing the phonon scattering (in addition to producing
tunneling centers).  First from XAFS, the off-center atom is partially bonded
to the nearest neighbor Ga/Ge atoms on the side of the cage\cite{Bridges03},
forming a random array of symmetry-breaking defects.  We show below that a high
density of such defects can lead to a $\kappa$ $\sim$ T dependence at low and
medium temperatures.  A likely more important consequence of this bonding is
that the motions of the off-center rattler atom are coupled more strongly to
just a few of the closest atoms in the surrounding Ga/Ge framework and hence to
the phonon modes - i.e.  the motions of the Eu2 and nearest Ga/Ge atoms are
positively correlated.

%Footnote: relatively low values for the broadening $\sigma$, of the Eu2-Ga/Ge pair
%distribution function below 40K from EXAFS experiments indicates that the
%motions of the Eu2 and nearest Ga/Ge atoms are positively correlated.

We first address the use of the tunneling model and the range of reasonable
tunneling splittings. Sethna and coworkers\cite{Grannan88}
suggest that for the light CN molecule the tunneling splittings (for
reorientations of the CN axis) might extend to about 1K; measured tunneling
splittings on similar, more dilute systems (CN$^-$ and OH$^-$) are less than
3K\cite{Bridges75,Narayanamurti70}.  For heavy off-center ions such as Ag$^+$
and Cu$^+$ in the alkali halides, the observed tunneling energies are also low,
much less than 1K\cite{Bridges75}; only for the light off-center $^7$Li$^+$ ion
does the tunneling splitting reach 1.2K at normal pressures\cite{Herendeen69}.
The tunneling behavior can be finely tuned by applying hydrostatic pressure to
reduce the potential barrier between equivalent off-center
sites\cite{Morgan87,Wang92}, but even in such cases the tunneling splittings
for heavy atoms such as  Ag$^+$ are about 1K when the off- to on-center
transition begins to occur\cite{Morgan87}.

Within the simple two-well tunneling model, having a tunneling splitting
($\sim$ 10-15K) that is $\sim$ 10\% of the attempt frequency (the rattler
frequency) is inconsistent with the assumptions for the model;  for a heavy
mass such as Eu, it leads to an unreasonably low potential barrier between the
wells ($<$ 1 meV (12K);  - i.e.  less than the tunneling splitting). On the
experimental side, a broad distribution of tunneling states should also lead to
a linear contribution to the heat capacity at low T when the tunneling states
are excited; however the heat capacity for the Ba, Sr, and Eu systems at low T
is exponential and can be fit to an Einstein model\cite{Sales01}; thus there is
no evidence for tunneling states extending up to 10-15K in such data.  With the
recent observations that the tunneling frequencies are at most
0.02K\cite{Hermann04} for the Eu system, another mechanism is needed
to describe $\kappa$ at low T.

Since atomic forces are generally a very strong function of the distance
between the atoms/ions, the force between the central atom (e.g. Sr2 or Eu2)
and each of the nearest equivalent Ga/Ge atoms in the cage, will increase
substantially for an off-center displacement. The resulting vibrations of the
off-center rattler/cage system then depend on the rigidity of the cage (the
Ga/Ge cage is quite stiff\cite{Chakoumakos00,Sales01}), the Eu2-Ga/Ge bond
strength, and the number of Eu2-Ga/Ge bonds. In the limit of an idealized rigid cage,
there would be no motion of the cage atoms (i.e. no phonon-coupling) as the rattler
atom vibrates; the reduced mass of the local mode would be that of the rattler
atom as assumed previously\cite{Chakoumakos00,Chakoumakos01,Bridges03}.
For the opposite extreme of a very soft cage, the local
mode reduced mass for motion along the bond direction would approach the
reduced mass of the atom-pair involved, e.g.  Eu and Ge in
Eu$_8$Ga$_{16}$Ge$_{30}$. In this case, there would be large motions  of
the nearest cage atoms (Ge) as a result of a local mode vibration, which would
couple directly to the phonons of the clathrate framework. Because the clathrate cage
is quite stiff, the reduced mass is expected to be close (but not equal) to the
free rattler mass.

To further understand this coupling, consider a 1-D model of a cage (mass $M$)
connected to a rattler (mass $m$), and let the respective displacements
be $x$ and $X$. Then $Mx = mX$; we
expect the matrix element for scattering to be proportional to $x/X$=$m/M$, and the
coupling to $(x/X)^2$.  The effective mass of the cage decreases when 
the rattler moves off-center (because only a small fraction of the cage atoms are directly
bonded to the rattler) and thus the coupling to phonons increases. Also note that
for the off-center case, rattler vibrations could be either $\sim$ radial (along the bond)
or perpendicular to the bond. The latter are similar to libration modes and would
have a lower $\Theta_E$ than for radial motion which was
not considered for the KCl:KCN system.

To investigate the various contributions to $\kappa$ we first
reproduced the calculation of Nolas {\it et al.}\cite{Nolas00}. $\kappa$ is
given by:

\begin{eqnarray}
\kappa & = & \frac{1}{3} \int_{0}^{\omega_D} c_s C(\omega,T) l(\omega,T) d\omega\\
l(\omega) & = & (\Lambda_R^{-1} + \Lambda_{res}^{-1} + \Lambda_{TS}^{-1})^{-1} + l_{min}\\
\label{kappa}
\Lambda_R^{-1} & = & D (\hbar \omega/k)^4 \nonumber\\
\Lambda_{res}^{-1} & = & \sum_{i} C_i \omega^2 T^2/[(\omega_{E_i}^2 - \omega^2)^2 -\gamma_{E_i} 
(\omega_{E_i}\omega)^2] \nonumber\\
\Lambda_{TS}^{-1} & = &  A(\hbar\omega/k_B)tanh(\hbar\omega/2k_BT) \nonumber\\
& &  +(A/2)(k_B/\hbar\omega + B^{-1}T^{-3})^{-1}
%\label{kappa}
\end{eqnarray}

\noindent where $c_s$ is the speed of sound, C($\omega$,T) is the heat capacity
of phonons with frequency $\omega$ and $l(\omega)$ is the total mean free path
which has three components - Rayleigh scattering ($\Lambda_R$), resonant
scattering from the Einstein modes ($\Lambda_{res}$) and a contribution from a
broad distribution of tunneling states ($\Lambda_{TS}$); the lower limit is
constrained to $l_{min}$.  The Einstein frequencies $\omega_{E_i}$ were
obtained from structural data and the constants A, B, C$_1$, C$_2$, D,
$\gamma_{E_i}$ and $l_{min}$ are given in ref.  \onlinecite{Nolas00}. We
obtained similar fits although the constants varied slightly.  We then reduced
the coupling to the resonant modes - the C$_i$ constants - by a factor of 10
and left the other terms unchanged.  This result is shown by the dashed line
(Fig. \ref{nolas_sales}a) and is very similar in shape to the data of
Sales\cite{Sales01} for Ba$_8$Ga$_{16}$Ge$_{30}$ (Fig \ref{nolas_sales}b).
Thus a major part of the difference in the T dependence of $\kappa $ between
Ba$_8$Ga$_{16}$Ge$_{30}$, and the Eu or Sr compounds can indeed be explained by
a decreased coupling between on-center Ba2 and phonons in the Ga/Ge framework as 
described above.

An important parameter for probing the rattler-phonon coupling is the reduced
mass $\mu$, of the local mode oscillator: however, $\mu$ is not an easy
parameter to obtain.  
%It requires a good measure of the rms vibration amplitude
%($\sigma$) toward the nearest cage neighbors, and zero-point motion must be
%observable.  There are at least three parameters that must be considered to
%model $\sigma^2$(T) - the Einstein temperature $\Theta_E$, $\mu$, and any small
%static contribution ($\sigma^2_{static}$) to the disorder at low T.  However in
%most experiments the signal-to-noise is not sufficient to determine 3
%independent parameters.  
One recent study of rattlers in the skutterudites\cite{Cao03}, found the reduced 
mass to be somewhat less than the atomic mass of the rattler (error $\sim$ 20\%), 
but more measurements are needed.

A second consequence of the rattler being partially bonded to a side of the
cage is the formation of a high-density, random network of symmetry-breaking
mass defects - i.e. one of the four off-center orientations is occupied, but
the off-center site varies randomly throughout the crystal.  The disorder
introduced by the random occupation of one of the off-center minima provides
insight as to why the crystal exhibits glass-like behavior.  This model has
similar features to the disorder introduced via irradiation of a quartz
crystal\cite{Berman50, Keppens96} but there is a key difference - the atoms are
randomly displaced in the irradiated sample which leads to a range of bond
lengths and hence to the possibility of a broad distribution of tunneling
states.  For the Eu/Sr clathrates the off-center displacement is well defined;
the disorder arises from a random occupation of the four off-center sites.
Locally such disorder changes the speed of sound as the phonon wave passes a
rattler atom; the resulting phase changes must then be included. Note that in
this case there is an interference in the scattering between different rattler
sites, whereas Rayleigh scattering assumes a single scattering process. 

Ziman has considered this type of scattering in some
detail\cite{Ziman60} - in his model, the effective mean free path for this
contribution is

\begin{equation}
\Lambda_{disorder}(q) = \frac{A}{L} \frac{c^2_s}{(\delta c_s q)^2}
\end{equation}

\noindent where $q$ is the wavenumber, $\delta c_s$ is the variation of the
sound speed, $L$ is the smallest distance between scatterers, and $A$ is a
numerical constant of order 1. For very short wavelengths $\lambda$ $<<$ $L$,
$\Lambda_{disorder}$ goes to a constant - but that is already included in the
model of Nolas {\it et al.}\cite{Nolas00} by the $l_{min}$ term. Using the
Debye model ($\omega = c_s q$) we then replaced the tunneling contribution
$\Lambda_{TS}^{-1}$ in Equ. \ref{kappa} by

\begin{equation}
\Lambda^{-1}_{disorder} = G \omega^2.
\end{equation}

In Fig. \ref{nolas_sales}a we show that with G$_{Eu}$  = 3.7 10$^{-19}$s$^2$/m
(G$_{Sr}$ = 6.4 10$^{-20}$s$^2$/m), and a slight change of the C$_i$ parameters
this model describes the T behavior of $\kappa$ (of Ref. \onlinecite{Nolas00}) just
as well as by using the tunneling model. Taking $L$ $\sim$ 5 10$^{-10}$m, these
numbers imply $\delta c_s$/$c_s$ $\sim$ 7\% (Eu) and 3\%(Sr).  Note that this
disorder mechanism is important at moderately long wavelengths - i.e. for
temperatures up to 20-30K, and alone would yield $\kappa$(T)$\sim$ T.

Attempts to fit the Sales {\it et al} data for Eu and Sr\cite{Sales01} using
the model of Nolas {\it etal}\cite{Nolas01} were not
successful\cite{Baumbach04} - the data from 2-10K vary nearly linearly with T,
and cannot be fit by the $\sim$ T$^2$ dependence for the tunneling model at low T.
However, we can fit the data quite well (Fig. \ref{nolas_sales}b) using
$\Lambda_{disorder}^{-1}$ but larger values for G and the C$_i$ are needed to fit
the lower thermal conductivity. 

For Ba, very much smaller values for C$_i$ are needed, which confirms the
earlier assertion that the coupling between phonons and the rattler vibrations
is reduced.  In addition, the $\Lambda_{disorder}^{-1}$ term cannot fit the low
T data for Ba below 12K; $\kappa$ is higher and the approximately $T^3$
dependence suggests that a finite sample size contribution is needed at low T
- details will be given in a separate paper\cite{Baumbach04}.  The lack of a
 $\Lambda_{disorder}^{-1}$ term is consistent with Ba being on-center, as in
that case the $\Lambda_{disorder}^{-1}$ term should disappear.

In summary, we have pointed out that an off-center rattler enhances the
scattering of phonons in two ways.  First, the off-center displacement
increases the coupling of the rattler motion to the phonons and hence increases
a dip in the thermal conductivity near $\sim$30K. This provides an explanation
for the the lack of a dip in $\kappa$(T)  for the on-center Ba system. Second,
the off-center atom on the side of the cage, introduces a quasi-random set of
``extra" atoms that will produce local changes in the sound velocity. This in
turn leads to a mean free path that varies as $\omega^{-2}$, which provides an
alternative explanation for the reduction of the thermal conductivity for 1K
$<$ T $<$ 20K.  It results in a linear T dependence in this range which fits
the data of Sales {\it etal}\cite{Sales01} well.

Several questions remain to be answered - for the clathrates, what are the
effective masses of the local Einstein modes and what are the actual tunneling
splittings (there will be 3 levels) in these systems? Can the Sr system be
moved on-center via high hydrostatic pressure and thus allow an investigation
of variations in the rattler-phonon coupling? Can these systems be optimized
for thermoelectric applications over a particular T-range, by varying the
off-center rattler's mass and/or Einstein temperatures? More generally does the
off-center displacement play a role in producing the plateau/dip regions in
other glassy systems?

\acknowledgments

We thank J. Rudnick and S. Shastry for helpful discussions. The work at UCSC
was supported in part by NSF grant DMR0071863.  The experiments were performed
at SSRL, which is operated by the DOE, Division of Chemical Sciences, and by
the NIH, Biomedical Resource Technology Program, Division of Research
Resources.

\bibliographystyle{prsty}
\bibliography{/home/users/bridges/bib/bibli.bib}

\end{document}